\pdfoutput=1

\documentclass[sigconf,natbib=true]{acmart}

\setcopyright{none}
\renewcommand\footnotetextcopyrightpermission[1]{}

\usepackage[T1]{fontenc}

\usepackage[hang,flushmargin]{footmisc}

\usepackage{url}            
\usepackage{booktabs}       
\usepackage{amsfonts}       
\usepackage{nicefrac}       
\usepackage{microtype}      
\usepackage{amsmath}
\usepackage{enumitem}
\usepackage{graphicx}       
\usepackage{caption}
\usepackage{subcaption}
\usepackage{threeparttable}
\usepackage{color, colortbl}
\usepackage{xcolor}

\newcommand{\ignore}[1]{}

\usepackage{threeparttable,booktabs}
\usepackage{multirow}
\usepackage{balance}
\usepackage{xspace}
\usepackage{balance}
\usepackage{mdframed}
\usepackage{makecell}  
\usepackage{array}     
\usepackage{pifont}
\usepackage[frozencache=true,cachedir=minted-cache]{minted}
\definecolor{LightGray}{gray}{0.9}
\usepackage{ragged2e}

\newcommand{\ragas}{RAGAs\xspace}

\newcommand{\umb}{UMBRELA\xspace}
\newcommand{\anu}{Auto\-Nuggetizer\xspace} 
\newcommand{\fsc}{FactScore\xspace}
\newcommand{\trecrag}{TREC 2024 RAG Track\xspace}

\newcommand{\autonuggets}{Auto\-Nuggets\xspace}
\newcommand{\autonuggetsedits}{Auto\-Nuggets\-+Edits\xspace}
\newcommand{\autoassign}{Auto\-Assign\xspace}

\newcommand{\manualassign}{Manual\-Assign\xspace}
\newcommand{\manualnuggets}{Manual\-Nuggets\xspace}

\title{The Great Nugget Recall: Automating Fact Extraction and RAG Evaluation with Large Language Models}

\author{Ronak Pradeep}
\affiliation{%
  \institution{University of Waterloo}
  \state{Ontario}
  \country{Canada}
}

\author{Nandan Thakur}
\affiliation{%
  \institution{University of Waterloo}
  \state{Ontario}
  \country{Canada}
}

\author{Shivani Upadhyay}
\affiliation{%
  \institution{University of Waterloo}
  \state{Ontario}
  \country{Canada}
}

\author{Daniel Campos}
\affiliation{%
  \institution{Snowflake}
  \city{San Mateo}
  \country{USA}
}

\author{Nick Craswell}
\affiliation{%
  \institution{Microsoft}
  \city{Seattle}
  \country{USA}
}

\author{Jimmy Lin}
\affiliation{%
  \institution{University of Waterloo}
  \state{Ontario}
  \country{Canada}
}

\begin{document}

\begin{abstract}
Large Language Models (LLMs) have significantly enhanced the capabilities of information access systems, especially with retrieval-augmented generation (RAG).
Nevertheless, the {\it evaluation} of RAG systems remains a barrier to continued progress, a challenge we tackle in this work by proposing an automatic evaluation framework that is validated against human annotations.
We believe that the nugget evaluation methodology provides a solid foundation for evaluating RAG systems. 
This approach, originally developed for the TREC Question Answering (QA) Track in 2003, evaluates systems based on atomic facts that should be present in good answers.
Our efforts focus on ``refactoring'' this methodology, where we describe the \anu framework that specifically applies LLMs to both automatically {\it create} nuggets and automatically {\it assign} nuggets to system answers.
In the context of the \trecrag, we calibrate a fully automatic approach against strategies where nuggets are created manually or semi-manually by human assessors and then assigned manually to system answers.
Based on results from a community-wide evaluation, we observe strong agreement at the run level between scores derived from  fully automatic nugget evaluation and human-based variants.
The agreement is stronger when individual framework components such as nugget assignment are automated independently. 
This suggests that our evaluation framework provides tradeoffs between effort and quality that can be used to guide the development of future RAG systems.
However, further research is necessary to refine our approach, particularly in establishing robust per-topic agreement to diagnose system failures effectively.
\end{abstract}

\renewcommand{\shortauthors}{}
\pagestyle{empty}

\maketitle

\section{Introduction}

This paper tackles the challenge of evaluating long-form responses from retrieval-augmented generation (RAG) systems built on top of large language models (LLMs), as part of the \trecrag.
There is, obviously, tremendous excitement and interest in RAG, but we feel that the {\it evaluation} of RAG output remains deficient from many perspectives.
Furthermore, the lack of standardized evaluations presents a barrier to continued progress in information access, and more broadly, NLP as well as AI.

Our central hypothesis is that the nugget evaluation methodology \citep{voorhees-2003-evaluating-answers,Voorhees_TREC2003} provides a solid foundation for evaluating RAG systems.
This methodology was originally articulated more than two decades ago in the context of the TREC Question Answering (QA) Track for evaluating answers to free-form ``definition questions''.
This matches our RAG setting, where for a given query, there are atomic facts (i.e., nuggets) from different documents that a system must synthesize into a fluent and cohesive natural language response.

Given this starting point, our efforts have focused on ``refactoring'' the original nugget evaluation methodology in light of LLMs.
Specifically, we leverage LLMs to both automatically {\it create} nuggets (nugget creation) and automatically {\it assign} nuggets (nugget assignment) to system-generated answers.
We implement our nugget evaluation methodology in the \anu framework.
While this is not the first attempt to automate nugget evaluations~\citep{Lin_Demner-Fushman_HLT-EMNLP2005,Lin_Demner-Fushman_IR2006}, the advent of LLMs provides opportunities that did not exist before.

We explored the following research questions:

\begin{itemize}

\item[{\bf RQ1}] For RAG systems, how well does our fully automatic nugget evaluation framework correlate with different manual nugget evaluation strategies (that vary in human involvement)?

\item[{\bf RQ2}] Does automating only nugget assignment result in stronger agreement with manual evaluations compared to fully automating the entire evaluation (including nugget creation)?

\item[{\bf RQ3}] Are there any noticeable differences between nugget assignments by humans versus LLMs?

\end{itemize}

\noindent In this work, we analyzed diverse runs from the \trecrag, a community-wide evaluation that attracted participation from dozens of teams around the world.
Our findings can be summarized as follows:

\begin{itemize}[leftmargin=*]

\item For \textbf{RQ1}, we find that scores from our fully automatic (end-to-end) nugget evaluation framework show strong correlations with manual nugget evaluations at the run level. This suggests that our approach can potentially serve as a good surrogate for manual evaluations in assessing RAG systems.

\item For \textbf{RQ2}, we find that automating only nugget assignment leads to stronger agreement with manual evaluations, compared to a fully automated evaluation where nuggets are constructed automatically.

\item For \textbf{RQ3}, our analyses suggest that LLM assessors appear to be more strict than NIST assessors in nugget assignment.
Additionally, the use of LLMs to provide ``draft'' nuggets in the nugget creation step does not appear to noticeably increase alignment with human nugget assignment.

\end{itemize}

\noindent The biggest advantage of our approach is minimal reinvention of the wheel, in that we can leverage the existing body of work that has gone into exploring the nugget evaluation methodology, e.g.,~\citep{lin-demner-fushman-2006-will,Lin_Zhang_SIGIR2007,Dang_Lin_ACL2007}.
For aspects that are not directly impacted by our use of LLMs, we can continue to assert findings from the literature without needing to carefully validate those claims again.
Furthermore, and unique to the TREC setup, we calibrate our fully automatic evaluation results against alternative strategies that involve NIST assessors to varying extents, representing different design choices in an overall evaluation framework (e.g., fully manual nugget creation vs.\ post-editing automatically generated nuggets).

The contribution of this work is, to our knowledge, the first large-scale study of automating the nugget evaluation methodology using LLMs for RAG systems. 
We demonstrate that system rankings derived from our fully automatic nugget evaluation process strongly correlate with those obtained from manual or semi-manual evaluations by human assessors.
This validation of automatic nugget creation and assignment represents an advance in the practical evaluation of RAG systems, offering a scalable and efficient alternative to labor-intensive manual assessments. 
Preliminary findings from our nugget evaluation were shared with the community in~\citet{Pradeep_etal_arXiv2024_RAG24}, but here we provide a much more thorough analysis of evaluation results.

It is important to emphasize that this work focuses exclusively on evaluating the recall of information nuggets within RAG answers. 
There are, of course, other important aspects of RAG evaluation, such as the assessment of citation support and the fluency of answers.
These are important to the overall evaluation picture, but beyond the scope of this paper, as it is impossible to tackle everything all at once.

\section{Background and Related Work}

\paragraph{RAG Systems and Benchmarks}

RAG in the context of LLMs has been used to generate entirely new, non-extractive answers to user queries \cite{guu_realm:2020,lewisrag,izacard-grave-2021-leveraging,borgeaud:2022}, where integrating external knowledge improves factual accuracy and mitigates issues such as hallucination~\cite{khandelwal:2020,gao-etal-2023-enabling}.
To better understand the nuances of these systems, RAG benchmarks have been developed to evaluate their effectiveness on different tasks, ranging from factoid QA~\cite{Yang2024CRAG, chenbench} to long-form answer generation~\cite{xu-etal-2023-critical,han-etal-2024-rag}.
Additionally, RAG benchmarks have been developed to explore various sources:\ knowledge graphs~\cite{pradeep-etal-2024-convkgyarn, feng2024cypherbench}, multilingual documents~\cite{thakur-etal-2024-knowing,thakur-mirage-bench}, coding~\cite{wang-coderag:2025, thakur-freshstack}, structured tabular data~\cite{ji2024target}, and images~\cite{hu2024mrag}.

\paragraph{LLM-as-a-Judge Evaluation.} 

Using LLM judges~\cite{ZhengLianmin:2306.05685:2023} in place of human judges has gained popularity for evaluating different aspects of RAG, such as factuality~\citep{krishna2024fact}, retrieval quality~\citep{salemi2024evaluatingretrievalqualityretrievalaugmented}, and robustness~\citep{sivasothy2024ragprobe} for specific aspects such as contextual faithfulness~\citep{nguyen2024sfr}. 
Researchers have also taken advantage of LLMs to directly perform pairwise comparisons between RAG responses~\cite{rackauckas2024evaluating,han-etal-2024-rag,Pradeep_etal_ECIR2025} and have suggested incorporating grading rubrics with strong LLMs~\citep{wang2024evaluating}.

Existing RAG frameworks assess responses along multiple dimensions.
For example, \ragas~\citep{es-etal-2024-ragas} introduced several metrics to evaluate faithfulness, answer relevance, and context relevance in RAG, leveraging LLMs (such as ChatGPT) for automatic evaluation based on custom prompts.
ARES~\citep{saad-falcon-etal-2024-ares} demonstrated that fine-tuning an open-source LLM with a small amount of training data suffices to produce an effective automatic judge.
Related, eRAG~\cite{salemi2024evaluatingretrievalqualityretrievalaugmented} utilizes the LLM in the RAG system itself to evaluate retrieval effectiveness.

\paragraph{Nugget Evaluation Methodology}
Introduced in the TREC QA Track in 2003~\citep{Voorhees_TREC2003}, the nugget evaluation methodology emphasized the identification of \emph{nuggets}, or essential facts relevant to an answer. 
\citet{voorhees-2003-evaluating-answers} demonstrated the importance of constructing nuggets to evaluate definition questions, thereby aligning complex question answering with retrieval-augmented generation. 
In both cases, queries require coherent answers drawn from information present in multiple relevant documents.

Further studies have refined and extended the basic nugget evaluation methodology.
For example, nuggets served as the basis for Summarization Content Units (SCUs)~\cite{nenkova-passonneau-2004-evaluating} to evaluate summaries in the Document Understanding Conferences.
Further refinements for question answering evaluation were introduced in later work~\cite{lin-demner-fushman-2006-will,Dang_Lin_ACL2007,Lin_Zhang_SIGIR2007}.
\citet{Lin_Demner-Fushman_HLT-EMNLP2005,Lin_Demner-Fushman_IR2006} explored automated approaches for nugget evaluation. 
\citet{nuggetir} introduced a nugget-based methodology for evaluating retrieval, addressing scalability and reusability issues inherent in traditional document-level relevance judgments.

Recently, the advent of LLMs has enabled researchers and practitioners to perform more reliable nugget evaluation automatically, an observation noted in concurrent papers~\citep{machine_gen, alaofi2024generative, thakur-freshstack}.
In this work, we incorporate these ideas into the \trecrag.

\paragraph{Related Nugget Evaluation Frameworks}
We can identify several attempts to build RAG evaluation frameworks around the notion of nuggets.
For example, \fsc~\citep{min-etal-2023-factscore} proposed to evaluate system-generated biographies from Wikipedia in terms of the percentage of atomic facts that are supported. 
Farzi and Dietz~\citep{farzi:2024} introduced the LLM-based RUBRIC framework, which evaluates the quality of system answers based on a rubric.
In a similar vein, \citet{arabzadeh2024comparison} used \emph{subtopics}, which are corpus-independent predicates generated by an LLM, to evaluate relevance, and \citet{machine_gen} used question and answer pairs as nuggets, manually identified by assessors.
Concurrently, \citet{thakur-freshstack} used an LLM to generate nuggets that decompose reference answers, which are then used to assess document relevance.
As most of these methods were primarily validated against traditional {\it ad hoc} retrieval tasks, it is unclear if they can be directly applied to RAG answer evaluation in our context.

\begin{table*}[!htbp]
\setlength{\tabcolsep}{3pt}
\centering
\resizebox{0.95\textwidth}{!}{
{\begin{tabular}{p{16cm}}
\toprule
{\bf Query}: how did african rulers contribute to the triangle trade \\[1ex]
{\bf Answer}: African rulers played a significant role in the triangular trade by capturing and supplying slaves to European traders. They actively participated in the trade by capturing individuals from inland areas and transporting them to the coast, where they were sold to European traders. $[$ \ldots{} $]$  In summary, African rulers contributed to the triangular trade by capturing and supplying slaves to European traders, driven by the economic benefits they received in exchange. Their involvement was crucial for the success and expansion of the transatlantic slave trade. \\
\bottomrule
\end{tabular}}}
\vspace{0.2cm}
\caption{Sample answer from the \trecrag for topic \texttt{2024-35227}. Note that here we have purposely omitted citations in the answer since this paper does not cover support evaluation.
The answer has been elided for length considerations.}
\label{tab:sample-sample}
\vspace{-0.5cm}
\end{table*}

In contrast, our nugget evaluation framework evaluates how many nuggets---gathered from a pool of relevant documents---are present in a system answer. 
Our experiments with \anu are grounded in the MS MARCO V2.1 segment collection, reflecting real-world search more accurately than (for example) Wikipedia alone (the target corpus of many other studies).
Crucially, our study includes correlations against high-quality human annotations from NIST assessors, similar to the established TREC QA Tracks, and features a ``fresh'' set of queries (reducing the risk of data pollution). 
The human annotations painstakingly gathered in the \trecrag provide valuable data for validating and calibrating automated LLM-based nugget creation and assignment.

\section{The Setup of the \trecrag}

The \trecrag comprises three distinct but inter-connected tasks:\ Retrieval (R), Augmented Generation (AG), and full Retrieval-Augmented Generation (RAG).
Participants were given 301 queries (topics);
the system task is to return, for the AG and RAG tasks, well-formed answers for each individual query (up to a maximum of 400 words).
The Retrieval (R) task can be viewed as an intermediate product in a full RAG pipeline.

Throughout this paper, we use topic \texttt{2024-35227} ``how did african rulers contribute to the triangle trade'' as a running example.
A system-generated answer is provided in Table~\ref{tab:sample-sample}; in this case, the answer was generated using GPT-4o~\citep{Pradeep_etal_ECIR2025}.
We have purposely omitted citations from this answer:\ 
actual submissions took the form of structured JSON data wherein systems explicitly attempted to link each answer sentence to passages in the corpus that (purportedly) provide evidence for the assertions made in the sentence.
This paper does {\it not} consider the evaluation of support, i.e., attempts to assess the veracity of the citations.

The Retrieval (R) task adopts a standard {\it ad hoc} retrieval setup, where systems are tasked with returning ranked lists of relevant passages from the MS MARCO V2.1 segment collection, comprised of 113 million passages~\cite{Pradeep_etal_ECIR2025, Upadhyay:2411.08275:2024,Pradeep_etal_arXiv2024_RAG24}.
The top-ranked passages retrieved for each query can then serve as grounding in the prompt input for LLM-based answer-generation systems.
In the Augmented Generation (AG) task, the organizers provided the participants with a fixed list of 100 retrieved results from which to generate their answers.
In the retrieval-augmented generation (RAG) task, participants were free to do whatever they wished in terms of retrieval from the corpus.
This paper focuses on analyzing the results of system-generated answers from only the AG and RAG tasks.

Our evaluation used queries (topics) sourced from Bing search logs, specifically selecting non-factoid queries that require comprehensive, multi-faceted, and potentially subjective responses~\citep{Rosset:2402.17896:2024}. 
The set of test queries was constructed near the submission deadline (late July 2024), rather than reusing queries that had already been publicly released in~\citet{Rosset:2402.17896:2024}. 
This strategy primarily addresses concerns about relevance judgment contamination, though we note that the corpus is likely to be already contained in LLM pretraining data given its web-based nature and open-source availability.
After additional curation by NIST, we arrived at the final test set of 301 queries.

\section{The Nugget Evaluation Methodology}
\label{section:nug-eva}

The nugget evaluation methodology comprises two main steps, which we paraphrase from \citet{Voorhees_TREC2003}, developed for the TREC 2003 QA Track:

\begin{itemize}[leftmargin=0.55cm]

\item[(1)] The assessor first creates a list of ``information nuggets'', where an information nugget is defined as a fact for which the assessor can make a binary decision as to whether a response contains the nugget.
At the end of this step, the assessor decides which nuggets are vital---nuggets that must appear in a response for that response to be good.

\item[(2)] The assessor then proceeds to the second step once the nugget list is created.
In this step, the assessor goes through each of the system answers and marks (whether or not) each nugget appears in the response.

\end{itemize}

\noindent Our \anu framework represents a ``refactoring'' of this decades-old nugget evaluation methodology, incorporating the latest advances in LLMs.
While there have been previous attempts to automate nugget evaluations~\citep{Lin_Demner-Fushman_HLT-EMNLP2005,Lin_Demner-Fushman_IR2006}, this effort represents the first since the advent of modern LLMs, which provide capabilities that did not previously exist.
In more detail, the nugget evaluation methodology comprises two main steps:

\paragraph{\bf Nugget Creation.}
This corresponds to the first step in the nugget evaluation methodology described by \citet{Voorhees_TREC2003}, often dubbed ``nuggetization''.
These nuggets capture elements of what should be in a good answer, divided into ``vital'' and ``okay'' categories.
``Vital'' nuggets are those, as \citet{Voorhees_TREC2003} articulates, must be present in a good response, while ``okay'' nuggets are ``good to have'', but are not absolutely necessary.

Nuggets must be created from an input set of documents under consideration; the input set is a design choice that we detail below.
In the original implementation of the evaluation methodology from 2003, NIST assessors manually formulated these nuggets based on documents in the pool that they assessed to be relevant.
That is, nugget creation followed relevance assessment (via pooling).
In our \anu framework, we have attempted to automate this entire process, using UMBRELA~\cite{Upadhyay_etal_arXiv2024} to generate relevance assessments for the pool and LLMs to generate what we call \autonuggets (see Section~\ref{section:auto-nuggets}).

It is important to note that while nuggets manifest as short natural language phrases or sentences, they are formulated at the semantic or conceptual level, and thus may or may not correspond to phrases or other lexical realizations in the sources.

\begin{table}[t]
\setlength{\tabcolsep}{3pt}

\renewcommand{\arraystretch}{1}
{\huge
\centering
\resizebox{0.46\textwidth}{!}{
{\begin{tabular}{p{16cm}}
\toprule
{\bf Automatic nugget creation (\autonuggets) using \umb qrels} \\
\midrule
            African rulers captured and sold slaves to Europeans (vital) \\
            African rulers waged wars to capture more slaves (vital) \\
            African rulers exchanged slaves for firearms (vital) \\
            African rulers' involvement was crucial for the trade's scale (vital) \\
            African rulers' trade led to increased internal slavery (okay) \\

\toprule
        {\bf Automatic nugget creation (\autonuggets) using NIST qrels} \\
\midrule
            African rulers captured and sold slaves to European traders (vital) \\
            African rulers exchanged slaves for firearms and goods (vital) \\
            African rulers' involvement was crucial for the transatlantic slave trade (vital) \\
            African rulers' cooperation enabled large-scale slave trade (vital) \\
            African rulers encouraged European traders to come to their ports (okay) \\
        \toprule
        {\bf Post-edited nuggets (\autonuggetsedits) using NIST qrels} \\
        \midrule
            African rulers sold slaves to European traders (vital) \\
            African rulers transported captives to coastal slave forts (vital) \\
            African rulers formed alliances with European traders (vital) \\
            African rulers' cooperation enabled large-scale slave trade (vital) \\
            African rulers' actions had a lasting negative impact on Africa (okay) \\
\bottomrule
\end{tabular}}}}
\vspace{0.2cm}
    \caption{Comparison of five sample nuggets for the query ``how did african rulers contribute to the triangle trade''\ from three conditions:\ \autonuggets using UMBRELA and NIST qrels, and \autonuggetsedits (i.e., after human post-editing).}
    \vspace{-0.25cm}
\label{tab:nuggets2}
\end{table}

\begin{table}[t]
\setlength{\tabcolsep}{2pt}
\small
\renewcommand{\arraystretch}{1}
{\huge
\centering
\resizebox{0.46\textwidth}{!}{
{\begin{tabular}{p{15cm}r}
\toprule
        \textbf{Nuggets} & \textbf{Assignment} \\
        \midrule
        {\bf Automatic nugget creation (\autonuggets) using \umb qrels} & {\bf \autoassign} \\ 
        African rulers captured and sold slaves to Europeans (vital) & Support \\
        African rulers waged wars to capture more slaves (vital) & Not Support \\
        African rulers exchanged slaves for firearms (vital) & Partial Support \\
        African rulers' involvement was crucial for the trade's scale (vital) & Support \\
        African rulers' trade led to increased internal slavery (okay) & Partial Support \\
        \midrule
        {\bf Post-edited nuggets (\autonuggetsedits) using NIST qrels} & {\bf \manualassign} \\
        African rulers sold slaves to European traders (vital) & Support \\
        African rulers sold war captives to European traders (okay) & Not Support \\
        African rulers sold criminals to European traders (okay) & Not Support \\
        African rulers sold debtors to European traders (okay) & Not Support \\
        African rulers' actions had a lasting negative impact on Africa (okay) & Not Support \\
        \bottomrule
    \end{tabular}}}}
    \vspace{0.2cm}
    \caption{Comparison of nugget assignment across different nugget creation approaches for the answer in Table~\ref{tab:sample-sample}.}
    \vspace{-0.25cm}
    \label{tab:assignment}
\end{table}

\paragraph{\bf Nugget Assignment.}
This corresponds to the second step in the nugget evaluation methodology described by \citet{Voorhees_TREC2003}.
After the nuggets have been created, the list can be treated like an answer key.
The assessor then reads the answer of each system to perform nugget assignment, which is a determination of whether each nugget appears in the response.
That is:\ Does a system answer contain this particular nugget?
In our \anu framework, nugget assignment is performed automatically by an LLM (see Section~\ref{section:auto-assign}).

It is important to note that the nugget assignment process is performed at the semantic or conceptual level, not merely based on lexical matching (and hence requires understanding, inference, and reasoning).
In particular, a nugget can be assigned to an answer (i.e., appears in the answer) even if there is no lexical overlap between the nugget itself and the system answer.

\medskip

\noindent
As a concrete example, the nuggets for topic \texttt{2024-35227} ``how did african rulers contribute to the triangle trade'' are shown in Table~\ref{tab:nuggets2} (top).
In this case, the nuggets were automatically created from documents considered by \umb to be at least related to the topic (more details in Section~\ref{section:auto-nuggets}).

After the nugget assignment process, we arrive at a record of which nuggets are found in which systems' answers.
For the sample answer in Table~\ref{tab:sample-sample}, the outcome of this is shown in Table~\ref{tab:assignment} (top).
At this point, it is straightforward to compute various metrics to quantify the quality of a response and the overall score of a run.
We refer interested readers to \citet{Voorhees_TREC2003} for details on how final scores were computed previously; we adopt a different approach to quantifying answer quality in our evaluation (see Section~\ref{scoring}).

\subsection{Nugget Creation}
\label{section:nuggets-creation}

\subsubsection{Automatic Nugget Creation}
\label{section:auto-nuggets}

\begin{figure}[t]
\begin{mdframed}[font=\scriptsize, roundcorner=10pt, linecolor=blue, linewidth=1pt, innerleftmargin=10pt, innerrightmargin=10pt, innertopmargin=10pt, innerbottommargin=10pt]
\textbf{SYSTEM:} You are NuggetizeLLM, an intelligent assistant that can update a list of atomic nuggets to best provide all the information required for the query.\\

\textbf{USER:} Update the list of atomic nuggets of information (1-12 words), if needed, so they best provide the information required for the query. Leverage only the initial list of nuggets (if exists) and the provided context (this is an iterative process).  Return only the final list of all nuggets in a Pythonic list format (even if no updates). Make sure there is no redundant information. Ensure the updated nugget list has at most 30 nuggets (can be less), keeping only the most vital ones. Order them in decreasing order of importance. Prefer nuggets that provide more interesting information.\\

Search Query: \{\texttt{\textbf{query}}\}\\

Context:\\

[1] \{\texttt{\textbf{seg}}$_1$\}

[2] \{\texttt{\textbf{seg}}$_2$\}

\ldots

[10] \{\texttt{\textbf{seg}}$_{10}$\} \\

Search Query: \{$\texttt{\textbf{query}}$\}\\

Initial Nugget List: \{\textbf{$n_{i-1}$}\}\\

Initial Nugget List Length: \{\textbf{\texttt{len}$(n_{i-1})$}\}\\

Only update the list of atomic nuggets (if needed, else return as is). Do not explain. Always answer in short nuggets (not questions). List in the form ["a", "b", ...] and a and b are strings with no mention of ". \\

Updated Nugget List: \\

\textbf{LLM:} [$n_{i, 1}$, \ldots]
\end{mdframed}
\vspace{-0.25cm}
\caption{Prompt for the iterative nuggetization at turn $i$.}
\label{fig:prompt.itnuggetizer}
\vspace{-0.25cm}
\end{figure}

The \anu framework begins by extracting a list of nuggets, or atomic information units, from a set of input documents.
This nugget creation process, dubbed ``nuggetization'', characterizes the information that should be contained in a high-quality answer to the user query.

To perform nuggetization, we employ GPT-4o through the Azure endpoint.
This process is run over all documents that are judged to be at least ``related'' to the query (grade $\geq$ 1).
Note that, depending on the actual evaluation condition (see below), these relevance judgments are either provided by NIST assessors (manual) or by \umb~\citep{Upadhyay_etal_arXiv2024} (automatic).
Details about relevance assessments are discussed in~\citet{Upadhyay:2411.08275:2024}.
Examples of these differences are shown in Table~\ref{tab:nuggets2}, comparing the top vs.\ the middle portions of the table.
There are minor differences, but overall the resulting nuggets are quite similar.

Using the prompt in Figure~\ref{fig:prompt.itnuggetizer},
the LLM iteratively updates a list of nuggets that collectively represent the key facts required to fully answer the query, conditioned on the provided contexts (passages). 
The first iteration for each query begins with an empty list.
Our prompt design encourages the model to produce comprehensive and diverse nuggets, ensuring broad coverage of different aspects of the user's information need.
This iterative approach allows for the generation of a rich set of nuggets, capturing both explicit and implicit information requirements derived from the query.
\anu aims to generate nuggets that are neither too broad nor too specific.
Informed by previous implementations of nugget evaluations, we limited generation to at most 30 nuggets.

\begin{figure}[t]
\begin{mdframed}[font=\scriptsize, roundcorner=10pt, linecolor=blue, linewidth=1pt, innerleftmargin=10pt, innerrightmargin=10pt, innertopmargin=10pt, innerbottommargin=10pt]
\textbf{SYSTEM:} You are NuggetizeScoreLLM, an intelligent assistant that can label a list of atomic nuggets based on their importance for a given search query.\\

\textbf{USER:} Based on the query, label each of the \{\textbf{\texttt{len}$(n_{f})$}\} nuggets either a vital or okay based on the following criteria. Vital nuggets represent concepts that must be present in a “good” answer; on the other hand, okay nuggets contribute worthwhile information about the target but are not essential. Return the list of labels in a Pythonic list format (type: List[str]). The list should be in the same order as the input nuggets. Make sure to provide a label for each nugget.\\

Search Query: \{\texttt{\textbf{query}}\}\\

Nugget List: \{\textbf{$n_{f}$}\}\\

Only return the list of labels (List[str]). Do not explain.\\

Labels:\\

\textbf{LLM:} ["vital", "okay", \ldots]
\end{mdframed}
\vspace{-0.25cm}
\caption{Prompt for determining the importance of nuggets. At each turn, at most 10 nuggets are passed to the LLM.}
\label{fig:prompt.nuggetscorer}
\vspace{-0.25cm}
\end{figure}

Once we have generated a set of nuggets for a query, the next step is to assess the importance of each nugget. 
We again use GPT-4o:\ the LLM is asked to assign an importance label of either ``vital'' or ``okay'' (see Section~\ref{section:nug-eva}) using the prompt shown in Figure~\ref{fig:prompt.nuggetscorer}.

At this point, we sort the nuggets in descending order of importance and select the first 20 nuggets.
This approach usually trims a few ``okay'' nuggets and, less frequently, some ``vital'' nuggets (when there are over 20 of them).
Note that these nuggets are ordered by decreasing importance, as imposed by the prompt in Figure~\ref{fig:prompt.itnuggetizer}.
The resulting nugget set, we dub \autonuggets.

\subsubsection{Semi-Manual Nugget Creation}
\label{sec:semi_manual_nuggets}

In the condition that we denote as \autonuggetsedits, NIST assessors post-edit nuggets that have been proposed by \anu as a ``rough draft''.
Here, we start with the set of documents that have been manually judged by NIST assessors to be at least ``related'' to the query (grade $\geq 1$), which serves as the input to \anu to create nuggets automatically (per above).
Note that in this case, the input set of documents has already been judged by humans to be at least related to the user's query, unlike with the \umb labels, which are generated automatically.

The automatically generated nuggets that we provide to NIST assessors are prepared in a slightly different way:\ for each query, we created a set of 30 nuggets without any indication of importance.
This over-generation is deliberate.
These nuggets are then edited by NIST assessors, who may add, eliminate, or combine nuggets.
The NIST assessors perform this task by concurrently considering the list of relevant documents for that topic.
Based on our logs, this process takes roughly an hour per topic on average, which suggests that the NIST assessors are not merely ``rubber stamping'' automatically generated nuggets, but are actually carefully deliberating over how they are formulated.

An example of this process is shown in Table~\ref{tab:nuggets2} for our running example.
On the top, we have the automatically generated nuggets using automatically generated \umb relevance judgments; in the middle, we have automatically generated nuggets using manually generated relevance labels; and on the bottom, we have post-edited nuggets generated using manually generated relevance labels.
Despite some differences, resulting nuggets from the different approaches are quite similar semantically.

\subsubsection{Manual Nugget Creation}
\label{sec:manual_nugget}

In the condition that we denote as \manualnuggets, NIST assessors examine the documents that have been manually judged to be at least ``related'' to the query (grade $\geq 1$) to create a set of at most 20 nuggets with their importance.
This followed the same procedure as the TREC 2003 QA Track; manual nugget creation is known to be a time-consuming and labor-intensive task.
Each topic required substantial assessor effort to thoroughly analyze the relevant passages and to synthesize the contents of those passages into a concise set of nuggets.
This took roughly 2.5 hours per topic, based on our measurements.

\subsection{Nugget Assignment}
\label{section:nuggets-assignment}

\begin{figure}[t]
\begin{mdframed}[font=\scriptsize, roundcorner=10pt, linecolor=blue, linewidth=1pt, innerleftmargin=10pt, innerrightmargin=10pt, innertopmargin=10pt, innerbottommargin=10pt]
\textbf{SYSTEM:} You are NuggetizeAssignerLLM, an intelligent assistant that can label a list of atomic nuggets based on if they are captured by a given passage.\\

\textbf{USER:} Based on the query and passage, label each of the \{\textbf{\texttt{len}$(n_{f})$}\} nuggets either as support, partial\_support, or not\_support using the following criteria. A nugget that is fully captured in the passage should be labeled as support. A nugget that is partially captured in the passage should be labeled as partial\_support. If the nugget is not captured at all, label it as not\_support. Return the list of labels in a Pythonic list format (type: List[str]). The list should be in the same order as the input nuggets. Make sure to provide a label for each nugget.\\

Search Query: \{\texttt{\textbf{query}}\}\\

Passage: \{\textbf{$p$}\}\\

Nugget List: \{\textbf{$n_{f}$}\}\\

Only return the list of labels (List[str]). Do not explain.\\

Labels:\\

\textbf{LLM:} ["support", "not\_support", "partial\_support", \ldots]
\end{mdframed}
\vspace{-0.25cm}
\caption{Prompt for nugget assignment. At each turn, at most 10 nuggets are passed to the LLM.}
\label{fig:prompt.nuggetassigner}
\vspace{-0.25cm}
\end{figure}

\subsubsection{Automatic Nugget Assignment}
\label{section:auto-assign}

The final component of our \anu framework, \autoassign, automatically assigns nuggets to systems' answers.
We adopt a listwise approach to nugget assignment, where the LLM is used to analyze an answer and determine if each nugget is fully supported (\texttt{support}), partially supported (\texttt{partial\_support}), or not supported (\texttt{not\_support}) by the answer.
Again, we employ GPT-4o through the Azure endpoint, with the prompt shown in Figure~\ref{fig:prompt.nuggetassigner}.
We iteratively prompt the model with at most 10 nuggets to evaluate assignment at each step.

\subsubsection{Manual Nugget Assignment}
\label{section:manual-assign}

In the manual nugget assignment process, we leveraged the expertise of the original NIST assessor involved in creating the list of nuggets (i.e., the same assessor performed nugget creation {\it and} nugget assignment). 
The assessor examines each answer, assigning each nugget one of three labels based on the extent of its support:\ \texttt{support}, \texttt{partial\_support}, and \texttt{not\_support} (same as with automatic assignment). 
It is worth clarifying that assessors in this step are not shown any LLM output, so this is a fully manual process.

By involving the same assessor responsible for nugget creation, we ensure consistency in nugget interpretation.
This continuity helps maintain reliability across the nugget assignment process, as the assessor applies the original context and intent behind each nugget to the assessment process.

An example is shown in Table~\ref{tab:assignment}:\ on the top, we show fully automatic assignment (\autoassign) of automatically created nuggets (\autonuggets) using \umb qrels; on the bottom, we show manual assignment (\manualassign) of post-edited nuggets (\autonuggetsedits) using NIST qrels.
The answer being evaluated is the one shown in Table~\ref{tab:sample-sample}.

\subsection{Evaluation Strategies}
\label{section:eval-combos}

Within our general \anu framework, we can instantiate a number of variants combining different design choices for nugget creation (Section~\ref{section:nuggets-creation}) and nugget assignment (Section~\ref{section:nuggets-assignment}).
In this work, we considered the following combinations:

\begin{enumerate}[leftmargin=0.5cm]

\item Automatic nugget creation with manual post-editing and manual nugget assignment (\autonuggetsedits {} / \manualassign).
Nuggets are initially generated automatically from passages judged relevant by NIST assessors and then refined by human assessors.
Nugget assignment is performed manually by NIST assessors.
This represents semi-manual nugget creation with manual assignment.

\item Manual nugget creation and assignment (\manualnuggets {} / \manualassign).
NIST assessors manually create nuggets from relevant passages and manually assign them to system answers. This is a fully manual evaluation baseline, mirroring the original TREC QA approach.

\item Fully automatic nugget creation and assignment (\autonuggets {} / \autoassign).
Nuggets are automatically generated from documents assessed relevant by \umb, and nugget assignment is performed automatically using \autoassign. 
This represents fully automated, end-to-end evaluation.

\item Semi-manual nugget creation with automatic nugget assignment.
In this condition, we use the nuggets from conditions (1) and (2), which involve human assessors in nugget creation (\autonuggetsedits and \manualnuggets, respectively).
However, we replace manual nugget assignment with automatic nugget assignment using \autoassign, i.e., \autonuggetsedits {} / \autoassign and \manualnuggets {} / \autoassign, respectively.
This condition isolates the impact of automating only the nugget assignment step.
\end{enumerate}

\subsection{Scoring}
\label{scoring}

At this point in the evaluation, we have already completed nugget creation and nugget assignment.
For each query, we have a list of nuggets, and for each system answer, we have a record of which nuggets it contains, in terms of a three-way judgment:\ \texttt{support}, \texttt{partial\_support}, and \texttt{not\_support} (either manual or automatic nugget assignment).

The final step is to compute the score for a system answer to a query $q$.
The score of a run is simply the mean of the score for each query.
We compute the following scores per query:

\paragraph{All Strict ($A_{\textrm{strict}}$)}
We calculate a score based on all nuggets in an answer, but with strict nugget matching.
For a given nugget $i$:
\begin{align}
    p_i &= \begin{cases}
        1 & \text{if assignment = \texttt{support}} \\
        0 & \text{otherwise}
    \end{cases}
\end{align}
The ``All Strict'' score is then calculated as:
$$A_{\textrm{strict}} = \frac{\sum_i p_i}{N_\text{nuggets}}$$

\paragraph{Vital Strict ($V_{\textrm{strict}}$)} 
This score, used as the primary evaluation metric, applies the same strict matching criterion as $p_i$ above, but only over the vital nuggets, $p^\textrm{v}_i$.
The score is calculated as:
$$V_{\textrm{strict}} = \frac{\sum_i p_i^\textrm{v}}{N_\text{vital nuggets}}$$

\noindent We emphasize that the scores presented here focus exclusively on evaluating the recall of nuggets in RAG answers. 
This analysis does not encompass other crucial aspects of RAG evaluation, such as the assessment of citation support or fluency, which are important but are outside the scope of this paper.

\begin{table*}[t]
\begin{small}
  \begin{tabular}{ll | c  c c | c c | r r r}
    \toprule
    & \multirow{2}{*}{\textbf{Condition}} & \multirow{2}{*}{\textbf{\# topics}} & \textbf{avg} & \textbf{avg nugget} & \multicolumn{2}{c|}{\textbf{\% nuggets}} & \multicolumn{3}{c}{\textbf{\% assigned}} \\
    &&& \textbf{nuggets} & \textbf{length} & \textbf{vital} & \textbf{okay} & \textbf{~ ~ NS} & \textbf{~ ~ PS} & \textbf{~ ~ S} \\
    \midrule
    \midrule

    (1a) & \autonuggetsedits {} / \manualassign & 36 & 14.1 & 7.0 & 61.1 & 38.9 & 57.9 & 6.6 & 35.6 \\
    (1b) & \manualnuggets {} / \manualassign & 20 & 13.7 & 8.5 & 59.0 & 41.0 & 62.9 & 3.5 & 33.6 \\
    \midrule
    (2a) & \autonuggets {} /  \autoassign (1a subset) & 36 & 18.2 & 6.8 & 66.2 & 33.8 & 50.9 & 22.0 & 27.0 \\
    (2b) & \autonuggets {} /  \autoassign (1b subset) & 20 & 18.7 & 7.2 & 70.3 & 29.7 & 47.1 & 24.9 & 28.0 \\
    \midrule
    (3a) & \autonuggetsedits {} / \autoassign & 36 & 14.1 & 7.0 & 61.1 & 38.9 & 54.1 & 18.7 & 27.2 \\
    (3b) & \manualnuggets {} / \autoassign & 20 & 13.7 & 8.5 & 59.0 & 41.0 & 57.8 & 17.6 & 24.6 \\
    \midrule
    (4) & \autonuggets {} /  \autoassign (Overall) & 301 & 18.7 & 6.8 & 72.5 & 27.5 & 50.9 & 23.6 & 25.5 \\
  \bottomrule
  \end{tabular}
  \end{small}
  \vspace{0.2cm}
  \caption{Descriptive statistics for nugget creation and assignment under different conditions. }
  \vspace{-0.5cm}
  \label{tab:nugget_stats}
\end{table*}

\section{Results}

\subsection{Descriptive Statistics}
\label{section:results:stats}

For the \trecrag, NIST received 93 runs from 20 groups for the RAG task and 53 runs from 11 groups for the AG task. 
Given resource constraints, the NIST annotators were able to evaluate only the two highest priority submissions from each group across the RAG and AG tasks, which translates into 31 runs from 18 groups for RAG and 14 runs from 9 groups for AG.

To be clear, our analyses are over 56 topics that have been fully judged across the 31 RAG runs and 14 AG runs discussed above.
Of these, 36 topics were under the \autonuggetsedits {} / \manualassign condition and 20 topics were under the \manualnuggets {} / \manualassign condition.
Descriptive statistics for nugget creation and assignment are presented in Table~\ref{tab:nugget_stats}, broken down by the various experimental conditions.
Naturally, since condition (3) in Section~\ref{section:eval-combos}, corresponding to row~(4) in the table, is fully automated, we were able to evaluate all topics, whereas for the other conditions we were constrained by available resources.

Examining the nugget characteristics, we observe that conditions employing \autonuggets, rows (2a), (2b), and (4), tend to generate a higher average number of nuggets per topic (around 18) compared to conditions using \autonuggetsedits or \manualnuggets (around 14).
Interestingly, the average nugget length remains relatively consistent across all conditions, hovering around 7--8 tokens, suggesting a stable granularity in nugget extraction regardless of the automation level in nugget creation. 
Furthermore, humans tend to be more conservative in assigning the label of ``vital'' importance, likely due to a stricter evaluation of what constitutes essential information within the context of a topic. 

Focusing on nugget assignment, a clear distinction emerges between manual and automatic assignment.
Conditions employing manual assignment, (1a) and (1b), exhibit a substantially lower percentage of nuggets categorized as ``Partial Support''.
In contrast, the automatic assignment conditions, especially the directly comparable ones, (3a) and (3b), see higher percentages. 
This shift suggests that automatic assignment tends to distribute some human-deemed ``Support'' assignments to ``Partial Support'', potentially indicating a different interpretation or application of support criteria compared to manual assignment.
However, this behavior could simply be the result of the current prompt that we are using.

\subsection{Fully Automatic Nugget Evaluation}
\label{section:results:rq1}

To address \textbf{RQ1}, we examine the correlation between our fully automatic nugget evaluation methodology and variants that require manual intervention to various degrees, as detailed in conditions (1) and (2) in Section~\ref{section:eval-combos}.
We reiterate that the \autonuggets {} / \autoassign condition is fully automatic, requiring no manual intervention (specifically, nuggetization is based on automatically generated relevance labels).

Here, we adopt the meta-evaluation methodology common in the information retrieval literature dating back several decades~\cite{Zobel98,Voorhees_SIGIR1998,Buckley00,Buckley04,Sanderson05,Harman_2011}.
We evaluate runs both with our fully automatic nugget evaluation methodology and with the other variants involving humans.
Rank correlations (in terms of Kendall's $\tau$) between the system orderings induced by the different approaches capture the quality of our automatic metric.
Our goal is to obtain high rank correlations against human-based metrics.

\begin{figure}[t]
        \centering
        \scalebox{0.45}{
            \begin{tabular}{cc}
            \includegraphics[width=0.49\textwidth]{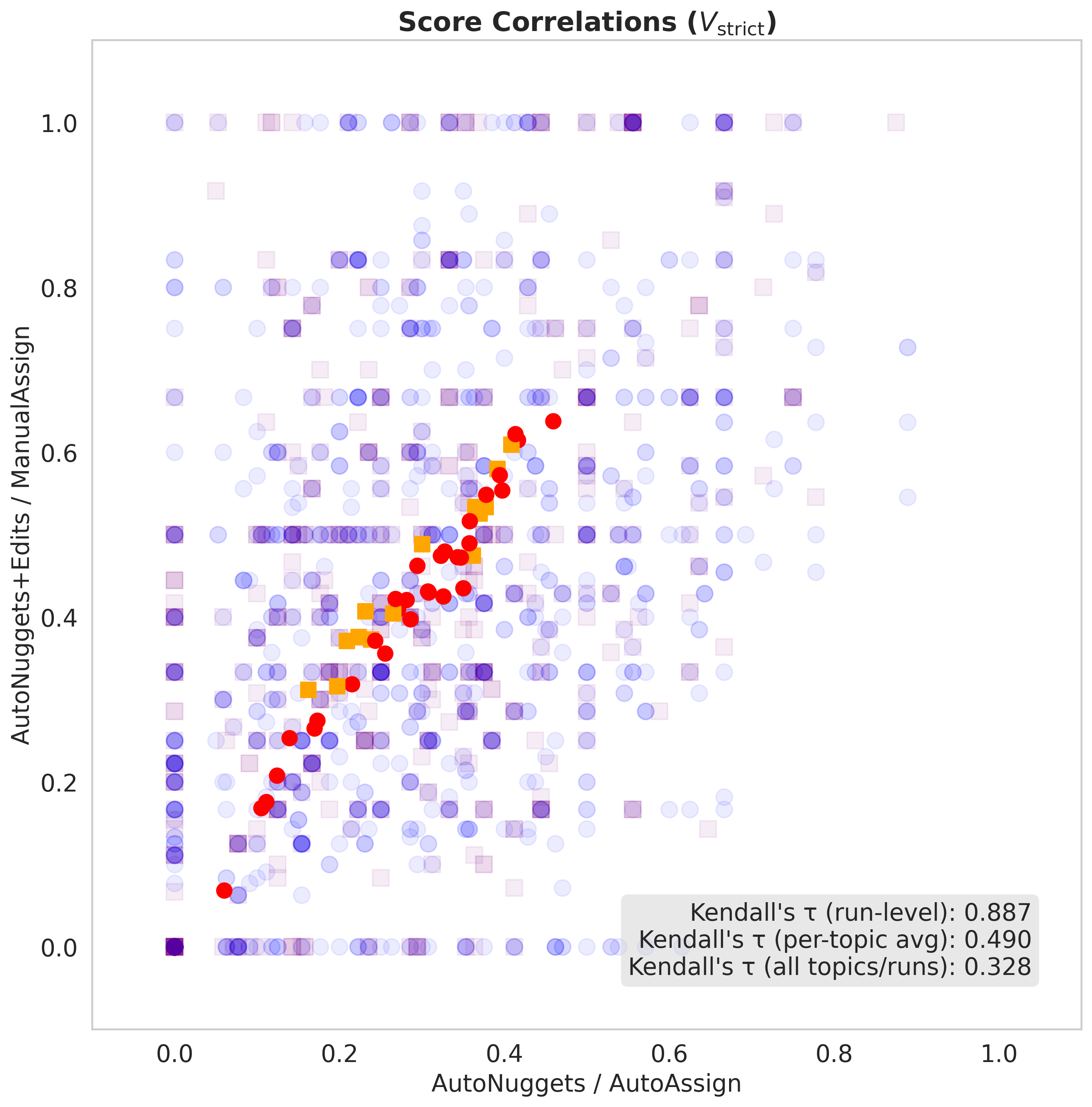} &
            \includegraphics[width=0.49\textwidth]{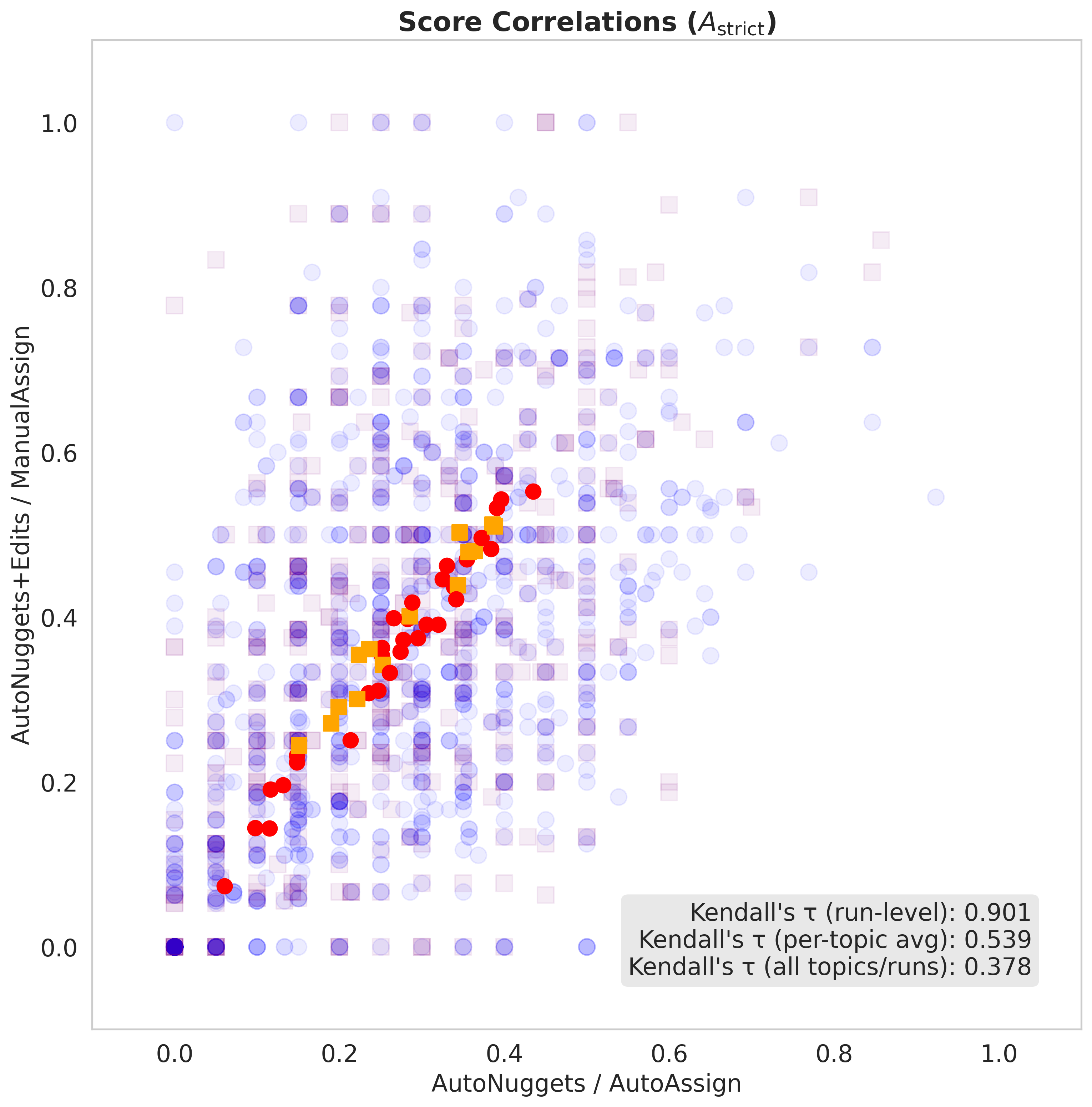} \\
            \includegraphics[width=0.49\textwidth]{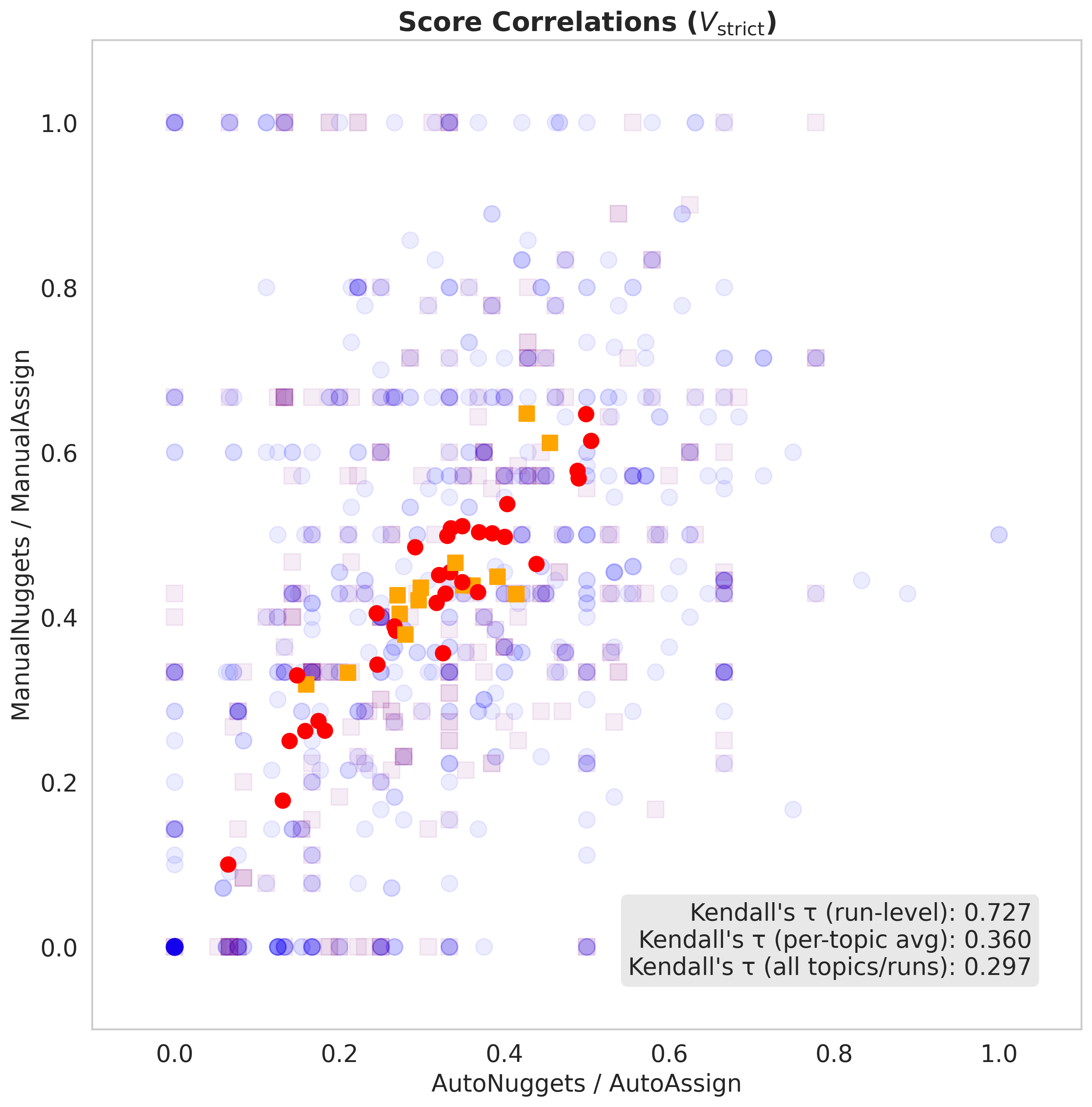} &
            \includegraphics[width=0.49\textwidth]{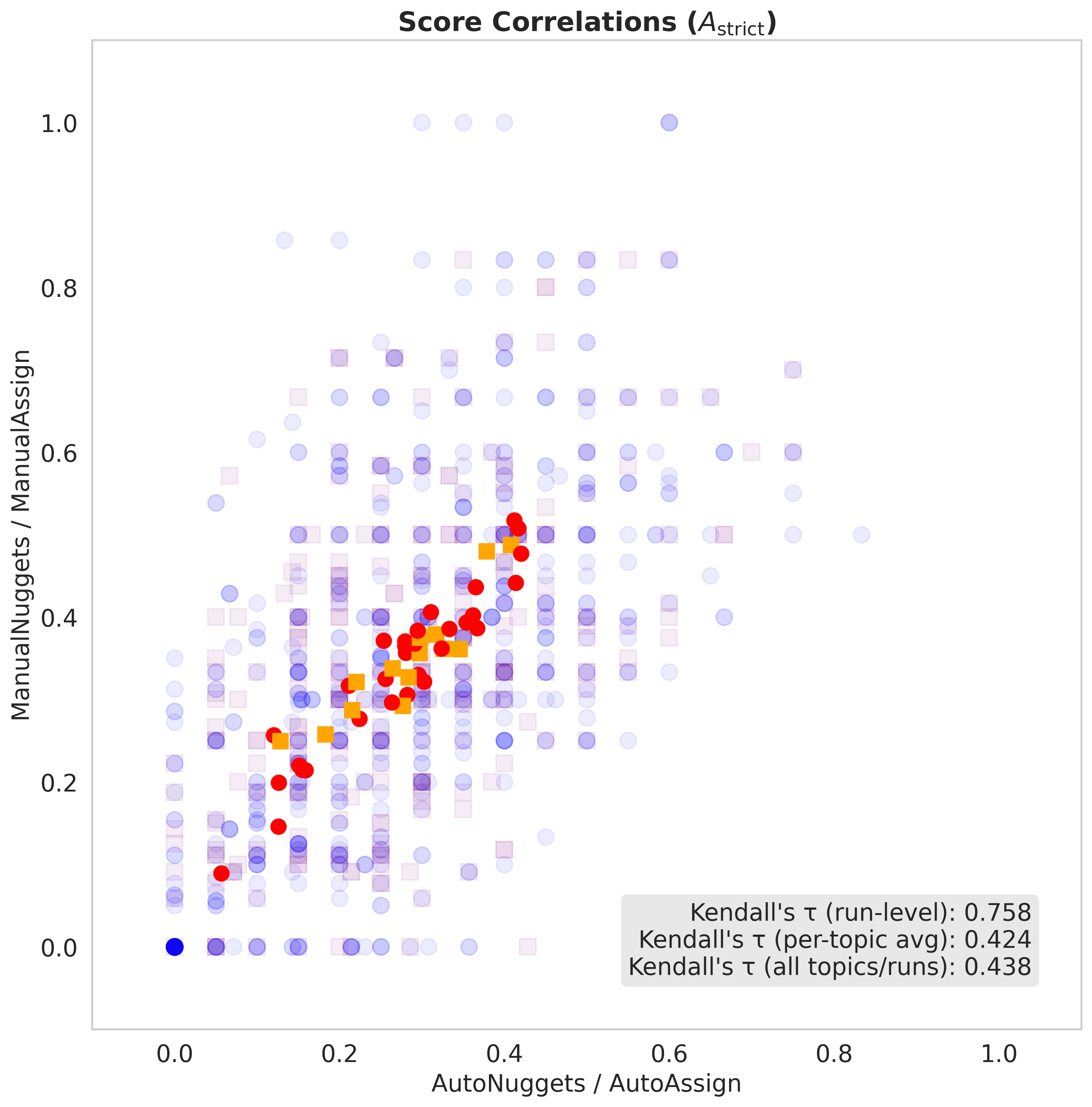}
            \end{tabular}
        }
    \caption{Scatter plots between manual vs.\ automatic $V_{\textrm{strict}}$ and $A_{\textrm{strict}}$ scores for AG and RAG runs.
    The $x$ axes show scores from \autonuggets {} / \autoassign and the $y$ axes show scores from different manual conditions.
    Red circles (RAG runs) and orange squares (AG runs) represent run-level scores. 
    Blue circles (RAG runs) and purple squares (AG runs) show all topic/run combinations. 
    The bottom-right box reports Kendall's $\tau$ correlations at the run level (red circles/orange squares), over all topic/run combinations (blue circles/purple squares), average of per-topic correlations.
    }
    \label{fig:manual_vs_auto}
    \vspace{-0.3cm}
\end{figure}

Figure~\ref{fig:manual_vs_auto} presents scatter plots of scores derived from the various experimental conditions.
In the left column, we show $V_{\textrm{strict}}$ and in the right column, $A_{\textrm{strict}}$.
In all these plots, the $x$-axes show the \autonuggets{} / \autoassign score.
The top row shows the comparison against the \autonuggetsedits {} / \manualassign condition, where nuggets are automatically generated and post-edited by humans, while assignment is manual.
The bottom row shows the comparison against the \manualnuggets {} / \manualassign condition, where both nugget creation and assignment are fully manual.
Red circles and orange squares represent run-level scores for RAG and AG tasks, respectively. 
Blue circles (RAG runs) and purple squares (AG runs) represent all topic/run combinations for the same.

We compute Kendall's $\tau$ in three different ways:\ (1) run-level correlations (i.e., we first compute run scores by averaging topic scores, and then compute correlations);
(2) average of per-topic correlations (per-topic avg), where we compute correlations per topic, and then average those;
(3) correlation across all topic/run combinations (all topics/runs), where each topic/run combination is considered an independent observation, and we compute Kendall's $\tau$ across all these observations.

For the \autonuggetsedits {} / \manualassign condition (top row), we observe a strong run-level Kendall's $\tau$ correlation of 0.887 for $V_{\textrm{strict}}$ and 0.901 for $A_{\textrm{strict}}$.
In IR meta-evaluations, these would be sufficient to ``validate'' a metric.
For the \manualnuggets {} / \manualassign condition (bottom row), the run-level Kendall's $\tau$ correlations drop to 0.727 for $V_{\textrm{strict}}$ and 0.758 for $A_{\textrm{strict}}$.
The correlations are slightly lower, but still good.
This can be partly explained by the much smaller topic count of 20:\ experimenting with a 20 topic subset of \autonuggetsedits results in correlations of 0.826 for $V_{\textrm{strict}}$ and 0.838 for $A_{\textrm{strict}}$.
This drop could perhaps be explained by fully manual nuggets ``looking different'' due to the assessors being anchored by the automatically generated nuggets (in the post-editing condition).
Nevertheless, results suggest good agreement between automatic and manual metrics at the run level.

However, examining the scatter of blue circles and purple squares representing individual topic/run combinations in all the scatter plots, we observe a greater dispersion compared to the run-level scores. 
The Kendall's $\tau$ correlation across all topic/run combinations is much lower, ranging from 0.297 to 0.438. 
The average per-topic Kendall's $\tau$ correlations fall to between 0.360 and 0.539. 
This indicates that while run-level rankings are well-preserved by our automatic evaluation, there is more variability at the topic level.

In summary, to answer \textbf{RQ1}:

\begin{itemize}
    \item[{\bf RQ1}] We find that scores from our fully automatic (end-to-end) nugget evaluation framework show strong correlations with manual nugget evaluations at the run level. This suggests that our approach can potentially serve as a good surrogate for manual evaluations in assessing RAG systems.
\end{itemize}

\noindent We observe stronger correlations with \autonuggetsedits than \manualnuggets, but Kendall's $\tau$ remains quite high even with fully manual nuggets.
We emphasize that while we achieve good {\it run-level} correlations, {\it per-topic} correlations are quite low, suggesting that our evaluation framework is inadequate for fine-grained debugging of individual answers.

\subsection{Automatic Nugget Assignment}
\label{section:results:rq2}

To investigate \textbf{RQ2}, we analyze whether automating nugget assignment in isolation, an expensive annotation step, leads to stronger agreement with manual evaluations compared to fully automating the entire framework.
That is, we hold the nuggetization approach constant, but vary the assignment method.

\begin{figure}[t]
        \centering
        \scalebox{0.45}{
            \begin{tabular}{cc}
            \includegraphics[width=0.49\textwidth]{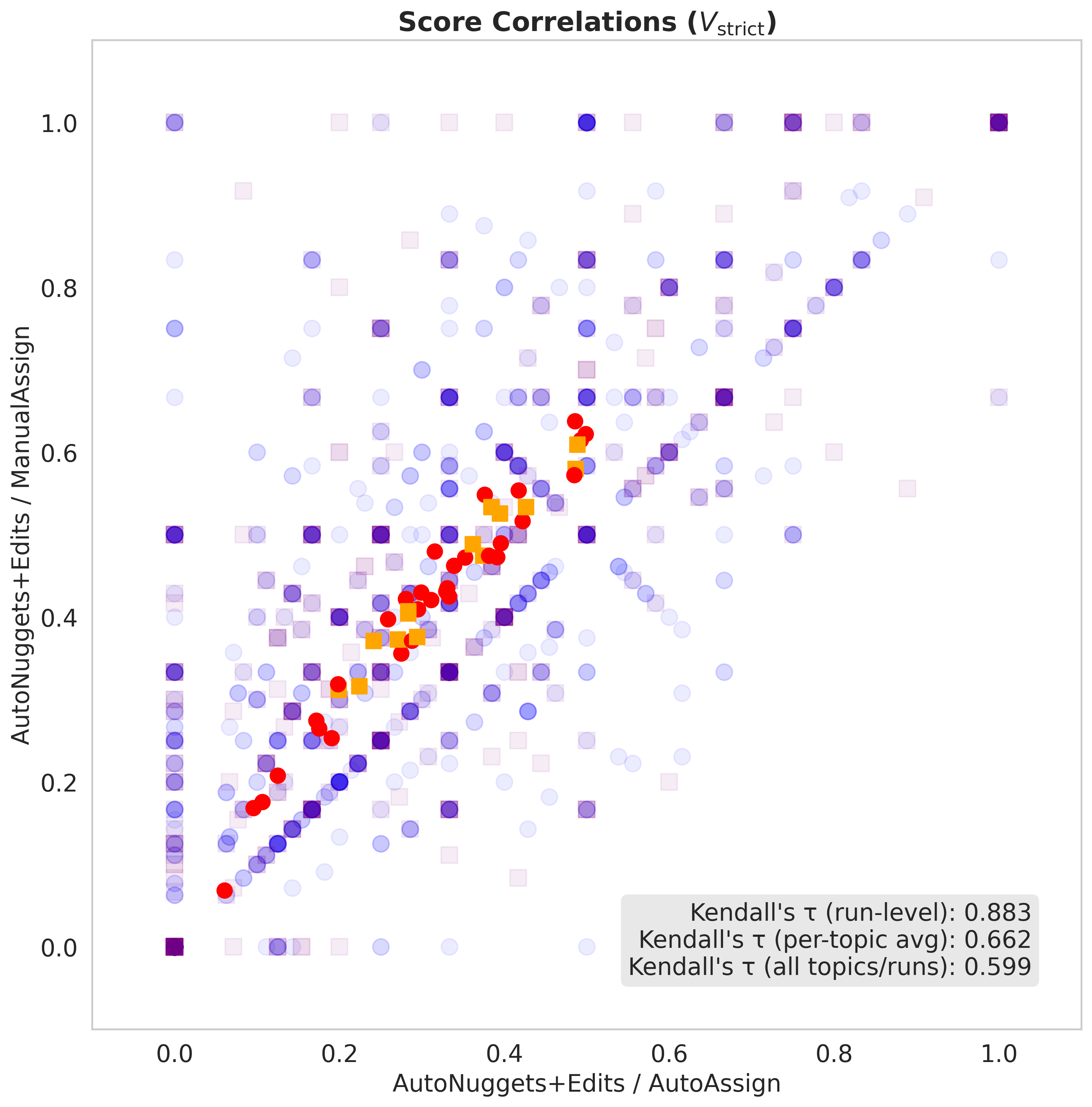} &
            \includegraphics[width=0.49\textwidth]{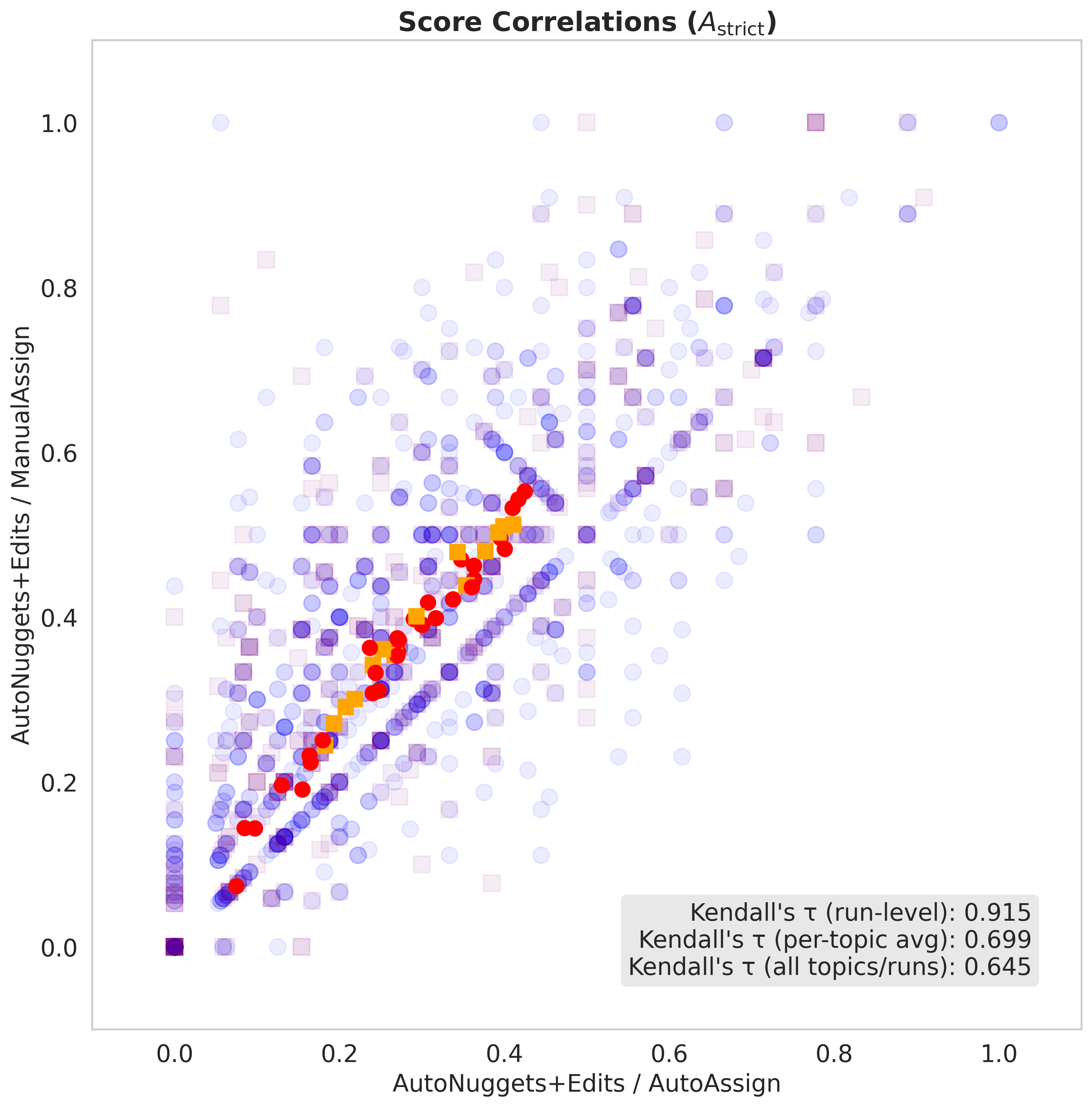} \\
            \includegraphics[width=0.49\textwidth]{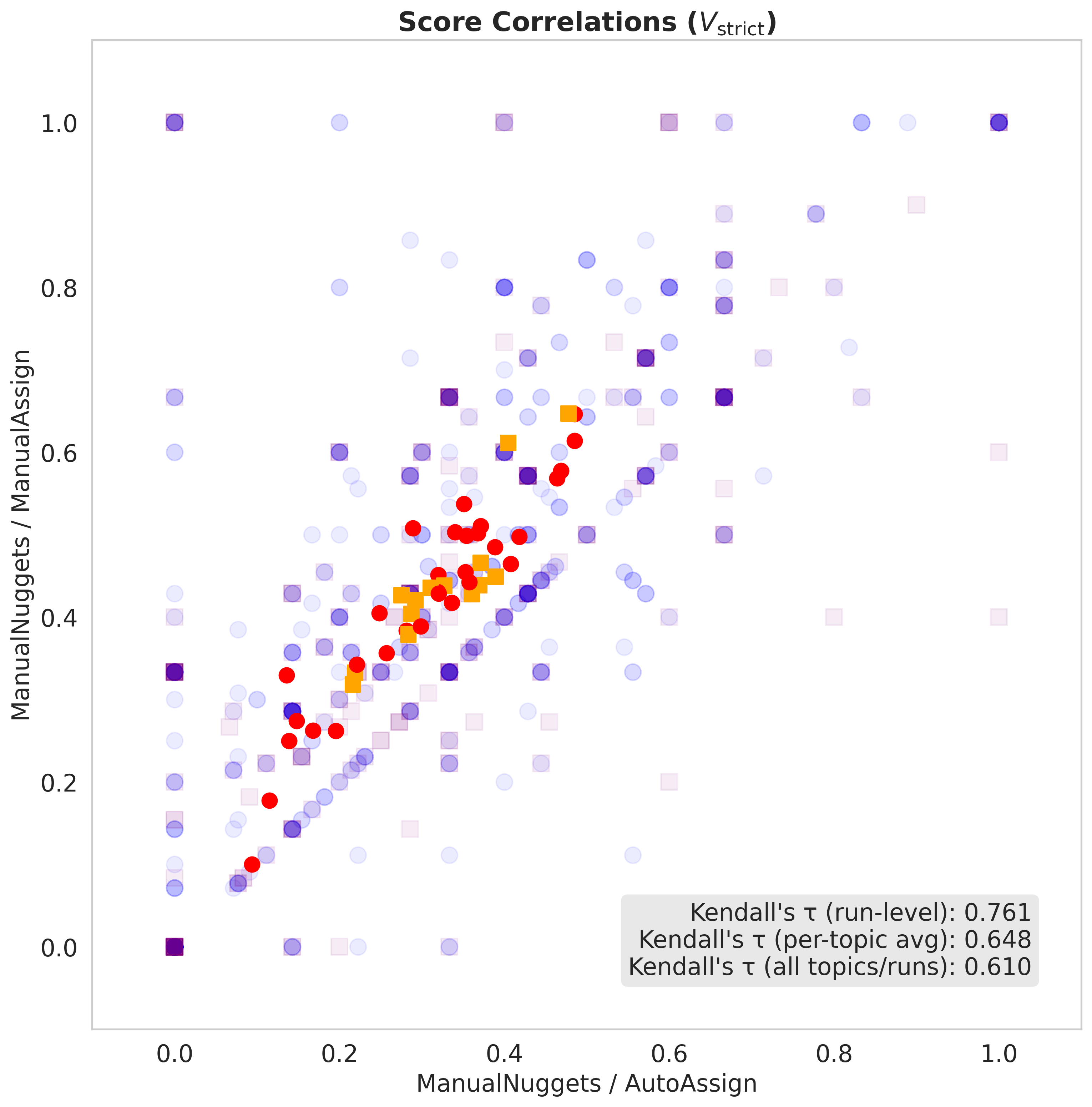} &
            \includegraphics[width=0.49\textwidth]{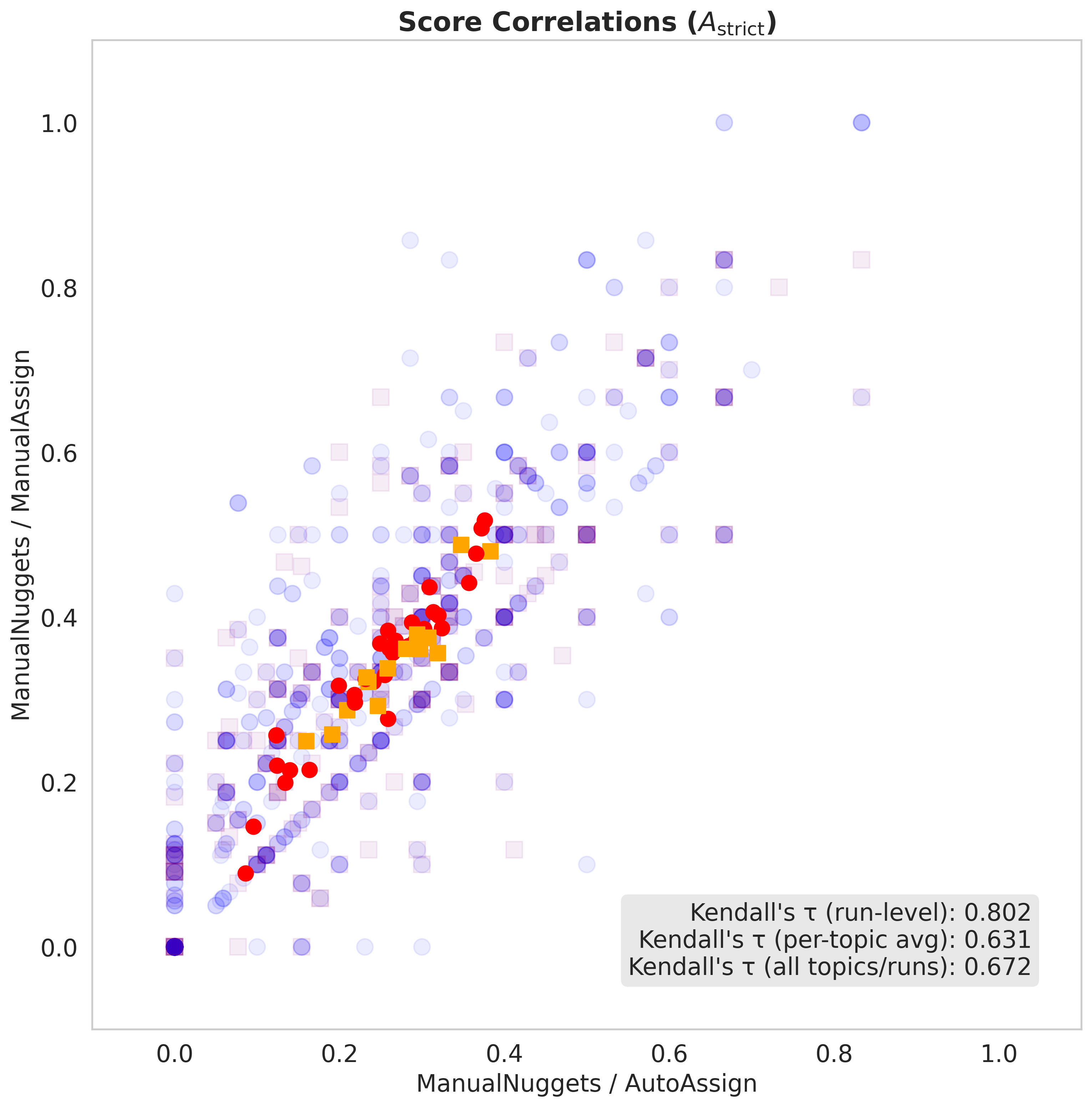}
            \end{tabular}
        }
    \caption{Scatter plots showing correlations designed to answer RQ2, isolating the effects of nugget assignment.
    Overall organization is identical to Figure~\ref{fig:manual_vs_auto}.
    }
    \label{fig:manual_vs_auto_assignment_only}
    \vspace{-0.2cm}
\end{figure}

Figure~\ref{fig:manual_vs_auto_assignment_only} compares scores with automatic vs.\ manual nugget assignment. 
The top row shows the \autonuggetsedits {} / \autoassign condition against \autonuggetsedits {} / \manualassign, isolating the effect of automating assignment when nuggets are post-edited by humans.
The bottom row compares \manualnuggets {} / \autoassign against \manualnuggets {} / \manualassign, isolating the effect of automating assignment when nuggets are manually created.

Comparing the top row of Figure~\ref{fig:manual_vs_auto_assignment_only} to the top row of Figure~\ref{fig:manual_vs_auto}, we observe that automating only nugget assignment (\autonuggetsedits {} / \autoassign vs.\ \autonuggetsedits {} / \manualassign) continues to show strong agreement at the run level. 
Additionally, it results in a higher per-topic average and all topics/runs Kendall's $\tau$ correlations for both $V_{\textrm{strict}}$ and $A_{\textrm{strict}}$.  
Specifically, for $V_{\textrm{strict}}$, the per-topic average Kendall's $\tau$ increases from 0.490 in the fully automatic (\autonuggets {} / \autoassign in Figure~\ref{fig:manual_vs_auto}) to 0.662 when only nugget assignment is automated (\autonuggetsedits {} / \autoassign in Figure~\ref{fig:manual_vs_auto_assignment_only}).  
Similarly, the all topics/runs Kendall's $\tau$ improves from 0.328 to 0.599.
For $A_{\textrm{strict}}$, we observe comparable improvements.
These metrics are arguably more sensitive to the nuances of topic-level evaluation, as they capture agreement across individual queries rather than just at the aggregate run level. 
The substantial increase in correlations suggests that automating nugget assignment while maintaining human-curated nuggets is perhaps a way forward to evaluate a large number of system responses, given that the assignment step occupies a lot of the assessment budget.

When we examine the bottom row of Figure~\ref{fig:manual_vs_auto_assignment_only}, where nuggets are manually created, we see a similar pattern of improvement.
Comparing \manualnuggets {} / \autoassign to \manualnuggets {} / \manualassign, we observe increases in per-topic average and all topics/runs Kendall's $\tau$ correlations for both $V_{\textrm{strict}}$ and $A_{\textrm{strict}}$. 
This further reinforces the finding that automating nugget assignment, regardless of whether nuggets are automatically generated and post-edited or manually created from scratch, leads to a better alignment with manual evaluations, particularly at the topic level.

To answer \textbf{RQ2}:
\begin{itemize}
    \item[{\bf RQ2}] We find that automating only nugget assignment leads to stronger agreement with manual evaluations, compared to a fully automated evaluation where nuggets are constructed automatically.
\end{itemize}

\noindent Our framework is not an ``all or nothing'' proposition.
Instead, there is a vast design space that supports different tradeoffs between effort and quality at the run- and topic-level.

\begin{figure}[t]
    \centering
    \scalebox{0.5}{
        \begin{tabular}{cc}
        \includegraphics[width=0.45\textwidth]{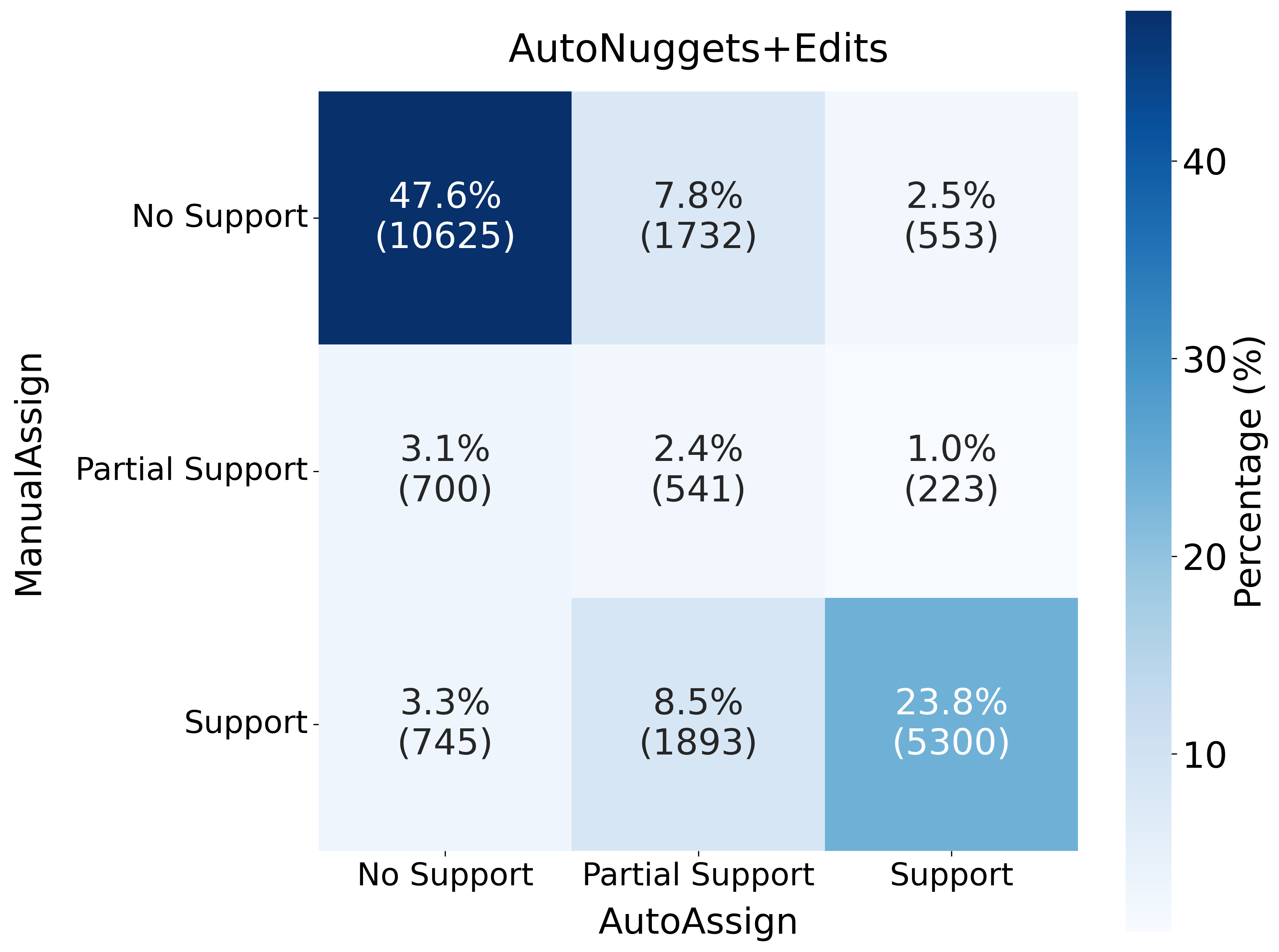} &
        \includegraphics[width=0.45\textwidth]{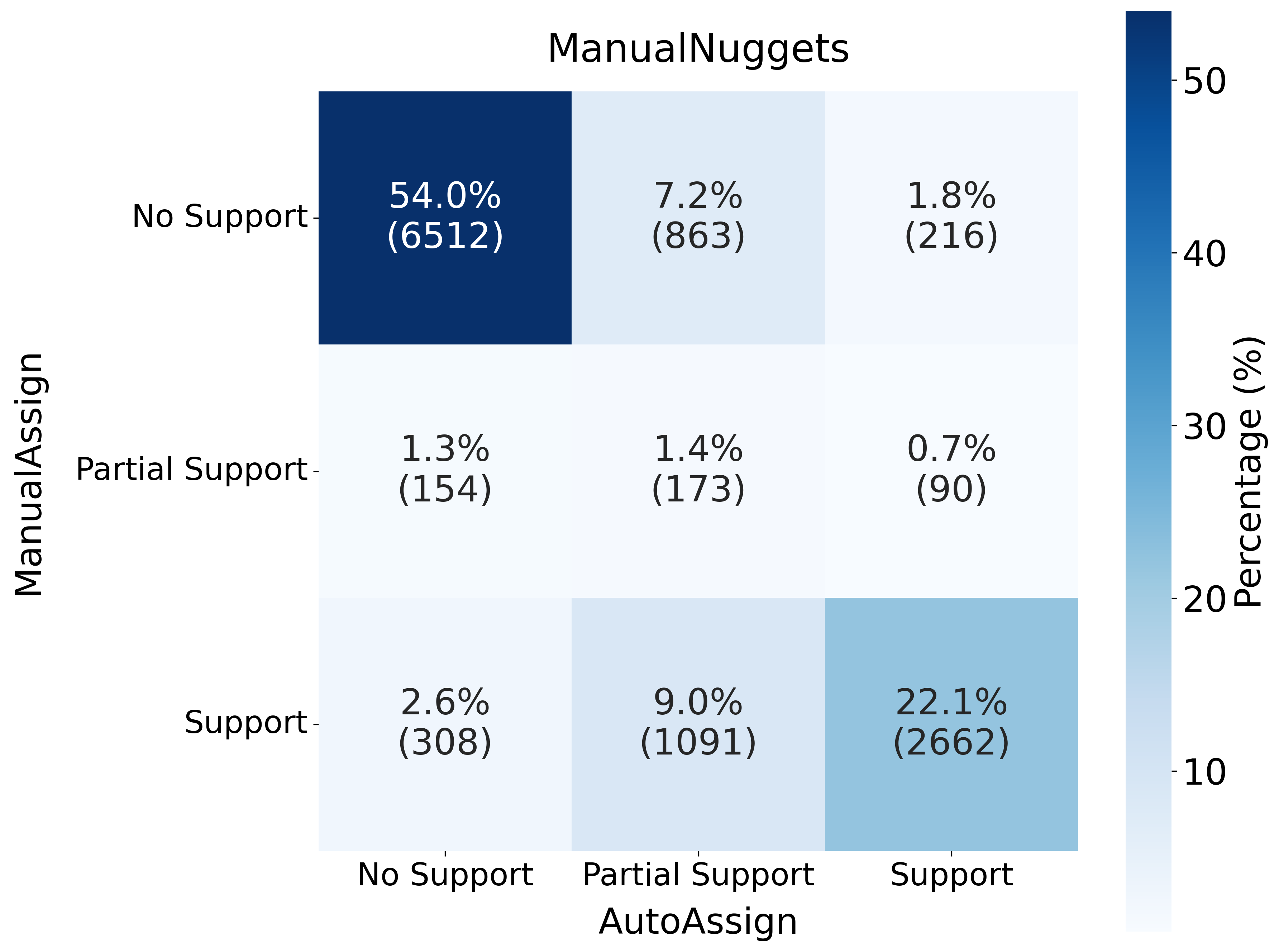} \\
        \end{tabular}
    }
    \caption{Confusion matrices showing agreement between automatic and manual nugget assignment strategies.}
    \label{fig:confusion_matrix_assignment}
    \vspace{-1em}
\end{figure}

\subsection{Nugget Assignment Confusion Matrices}
\label{section:results:rq3}

To address \textbf{RQ3}, we investigate quantitative differences between human and LLM nugget assignment. Figure~\ref{fig:confusion_matrix_assignment} presents confusion matrices that quantify agreements (and disagreements) in nugget assignment labels between automatic (\autoassign) and manual (\manualassign) approaches.
The left matrix compares \autoassign against \manualassign when using \autonuggetsedits nuggets, and the right matrix compares them when using \manualnuggets.  
Each matrix shows the percentages and counts of nugget assignments falling into different categories.

Examining the diagonal elements of both confusion matrices, we observe substantial agreement between automatic and manual nugget assignment, particularly for the ``Support'' and ``No Support'' labels.
For \autonuggetsedits nuggets (left matrix), 47.6\% of the (response, nugget) pairs resulted in both \autoassign and \manualassign labeling a ``No Support'' and 23.8\% a ``Support''. 
For \manualnuggets (right matrix), we see 54.0\% of the distribution with ``No Support'' and 22.1\% with ``Support''.
Partial support appears to be a source of disagreement in both conditions, as evidenced by only a small mass of the distribution contained in the central element of the matrices.

Off-diagonal elements reveal disagreements. 
In both matrices, we observe a tendency for \autoassign to assign ``Partial Support'' more frequently than \manualassign.
In general, we see a slightly larger mass of the distribution in the lower triangle part of both matrices compared to the upper triangle, indicating that LLM assessors are a bit stricter compared to NIST assessors. 

The use of LLMs to provide ``draft'' nuggets in the prior nuggetization step (\autonuggetsedits vs. \manualnuggets) does not appear to drastically alter the alignment across labeling in nugget assignment, as the confusion matrices for both nugget types exhibit similar patterns of agreement and disagreement.

To answer \textbf{RQ3}:

\begin{itemize}
    \item[{\bf RQ3}] Our analyses suggest that LLM assessors appear to be more strict than NIST assessors in nugget assignment.
Additionally, the use of LLMs to provide ``draft'' nuggets in the nugget creation step does not appear to noticeably increase alignment with human nugget assignment.
\end{itemize}

\noindent Two points of caution here, though:\
These findings are specific to the current implementation.
For example, it is entirely possible that a different LLM or a different prompt might elicit different behavior.
Another gap in our current understanding is the relationship between inter-annotator agreement and LLM--human differences.
This analysis only compares {\it one} set of nugget assignments between an assessor and a particular LLM implementation.
We do not know the variations exhibited by different human assessors and different LLMs, and how they compare, which would be an interesting question to tackle in the future.

\section{Conclusions}

While we are certainly not the first to propose a RAG evaluation methodology, we view our efforts as having two main distinguishing characteristics:\
First, by building on the nugget evaluation methodology dating back over two decades, we minimize reinvention of the wheel.
The information retrieval literature has a long tradition of deliberate and careful meta-evaluations that validate evaluation methodologies, and much work has examined different aspects of nugget evaluations.
For aspects of the evaluation that are not dependent on LLMs, we can simply build on existing findings.
Second, we have demonstrated that our \anu framework, which leverages LLMs for automatic nugget creation and assignment, achieves strong run-level agreement with manual and semi-manual evaluations, suggesting its viability as a scalable alternative for assessing RAG systems.

Under the \autonuggets {} / \autoassign condition in our \anu framework, we are able to provide evaluation scores for \emph{all} runs submitted to the \trecrag, across all 301 topics---at only the cost of LLM inference, while doing the same with human assessors remains prohibitively expensive.
Our insights reveal that automating nugget assignment in isolation yields stronger agreement, especially at the per-topic level, with manual evaluations than with full framework automation, highlighting the potential of a hybrid approach. 
Furthermore, we find that LLMs apply stricter assignment labels than human assessors, indicating the need for calibration. 
Our framework has the potential to enable rapid iteration on improving RAG systems with varying degrees of human involvement, while providing some confidence that the automatically generated metrics have some degree of correlation to answer quality as determined by human assessors.
For RAG, we can now potentially climb hills quickly and in a meaningful way!

\section*{Acknowledgments}

This work would not have been possible without the annotator team at NIST.
This research was supported in part by the Natural Sciences and Engineering Research Council (NSERC) of Canada.
Additional funding is provided by Snowflake, Microsoft via the Accelerating Foundation Models Research program, and an Institute of Information \& Communications Technology Planning \& Evaluation (IITP) grant funded by the Korean Government (MSIT)\ (No.\ RS-2024-00457882, National AI Research Lab Project).
Thanks to Corby Rosset for providing the test queries, based on the methodology developed in Researchy Questions \cite{Rosset:2402.17896:2024}.

\bibliographystyle{ACM-Reference-Format}
\bibliography{ref}

\end{document}